\documentclass[10pt]{article}

\usepackage{amsmath}
\usepackage{amssymb}

\usepackage{graphicx}

\usepackage{cite}

\usepackage{color} 

\usepackage[labelfont=bf,labelsep=period,justification=raggedright]{caption}

\makeatletter
\renewcommand{\@biblabel}[1]{\quad#1.}
\makeatother

\date{}

\begin{document}

\begin{flushleft}
{\Large
\textbf{IMDB network revisited: unveiling fractal and modular properties from a typical small-world network}
}
\\
Lazaros K. Gallos$^{1}$, 
Fabricio Q. Potiguar$^{2}$, 
Jos\'e S. Andrade Jr$^{3}$,
Hernan A. Makse$^{1,\ast}$
\\
\bf{1}  Levich Institute and Physics Department, City College of New York, New York, NY, US
\\
\bf{2} Faculdade de F{\'\i}sica, ICEN, Universidade Federal do Par\'a, Bel\'em, Par\'a, Brasil
\\
\bf{3} Departamento de F{\'\i}sica, Universidade Federal do Cear\'a, Fortaleza, Cear\'a, Brasil
\\
$\ast$ E-mail: hmakse@lev.ccny.cuny.edu
\end{flushleft}

\section*{Abstract}
We study a subset of the movie collaboration network, imdb.com, where only adult movies
are included. We show that there are many benefits in using such a network,
which can serve as a prototype for studying social interactions.
We find that the strength of links, i.e., how many times two actors have
collaborated with each other, is an important factor that can significantly 
influence the network topology. We see that when we link all actors in the 
same movie with each other, the network becomes small-world, lacking a 
proper modular structure. On the other hand, by imposing a threshold on the 
minimum number of links two actors should have to be in our studied subset, 
the network topology becomes naturally fractal. This occurs due to a large 
number of meaningless links, namely, links connecting actors that did not 
actually interact. We focus our analysis on the 
fractal and modular properties of this resulting network, and show that the 
renormalization group analysis can characterize the self-similar structure 
of these networks.

\section*{Introduction}
The study of real-life systems as complex networks \cite{albert02,boccaletti06} 
has allowed us to probe into large-scale properties of databases that relate to
a wide range of disciplines. Among them, social networks 
\cite{sbfn,bernard88,vanraan90,sna} play a special role
because we can follow the trails that are left behind social interactions 
(especially in their online form). Moreover, understanding the 
structure and the mechanisms behind the resulting web of those interactions
poses an extremely challenging problem with important implications for 
everyday life, such as disease spreading patterns 
\cite{sattenspiel88,kretzschmar96}.
In recent years, the development of databases \cite{imdb,acidb,impdb,dipdb} 
that depict such interactions 
has resulted in many networks that model different aspects of social systems. 
In fact, social networks can also be classified as collaboration networks 
because people belonging to them are usually indirectly connected 
through their common collaboration entity, be it scientific collaboration 
\cite{newman01}, company board membership \cite{davis97} or movie acting 
\cite{watts98}. 
Usually, these networks are bipartite: there are two types of nodes and links 
run only among unlike vertices. However, they can also be taken as a one-mode 
projection of the bipartite structure, connecting nodes of the same type 
whether they share a node in the bipartite structure \cite{newman01-02}.

The majority of social networks has a small-world character 
\cite{watts98,swnt}, substantiated by a small average short-path length 
and a large average clustering coefficient.
At the same time, many social networks have been shown to be fractal 
\cite{song05,song06}, i.e., they present a self-similar
character under different length scales. These properties can be seen 
by the optimal covering of a network with boxes of maximum diameter 
less than $\ell_B$ \cite{song07}, and a subsequent power-law scaling of the
required minimum number of boxes $N_B(\ell_B)\sim\ell_B^{-d_B}$, where $d_B$ is 
the fractal dimension of the network. 
Fractality in networks \cite{hartwell99,ravasz02,gallos07,galvao10,gallos12} 
is not a trivial property and in many cases it can be masked under the 
addition of shortcuts or long-range links on top of a purely fractal 
structure.

Recently, we have calculated analytically the small-world to large-world,
a self-similar structure with a few shortcuts, transition using renormalization 
group (RG) theory \cite{rozenfeld10}.
The basic idea of the RG method is to compute the difference between the average 
degree of the renormalized network,
$z_B$, and the average degree of the original network (the network without short-cuts), $z_0$,
for different values of $\ell_B$. 
The resulting relation is 
\begin{equation}
\label{rg_eq}
z_B-z_0 \sim m_B^\lambda, 
\end{equation}
where $m_B=N_B/N$ is the average number of nodes in a box of diameter $\ell_B$ 
(the average `mass' of the box). The exponent $\lambda$ 
characterizes the extent of the small-world effect. The $\lambda>0$ case corresponds to
a large number of shortcuts, so that successive renormalization steps lead to an increase
of the average degree by bringing the network closer to a fully-connected graph 
(small-world). When $\lambda<0$, only short-range shortcuts exist, so that the 
distance between nodes remains significant during successive steps and the network 
fractality is preserved (large-world).

In this work, we present an example of how to unmask fractality in a 
small-world network and elaborate on the methods that highlight the fractal 
characteristics. Our primal SW network is the well known IMDB movie 
co-appearance network \cite{imdb,watts98,rozenfeld10,barabasi99,amaral00}, where two actors 
are connected if they have participated in the same movie. This 
environment is very dynamic and actors participate in 
new movies continuously, creating many links and rendering the
network a typical example of a small-world prototype. On the other hand, the 
construction method of the IMDB dataset (each movie leads to
a fully connected subgraph and actors tend to specialize in one particular 
genre) seems to naturally lead to a fractal and modular structure.

In order to observe this transition more clearly, we focus our analysis on a 
smaller subset of the IMDB 
database, by taking into account movies that have been labeled by IMDB as adult,
and considering only the actors that have participated in them.
Except for the smaller and more easily amenable size, there are many
advantages in this choice: a) First, this network is a largely isolated subset
of the original actors collaboration network, since there are very few links
that connect actors in this genre with actors in other genres, so that it can 
safely be studied as a separate network on its own. b) Practically all these 
movies have been produced during the last few decades, so that the network is 
more focused in time, which reduces the inherent noise arising from 
connections between very distant actors. There are a number of examples of 
actors that had long lived carrers and are certainly connected to other 
actors that did not live in the same time period.

Only taking the adult subset of IMDB is not enough to observe fractality, since 
both sets are typical small-worlds. This observation is achieved when we consider 
the interaction strength between actors, $w_M$. This strength is 
defined as the number of times two actors were cast together in a movie. 
To consider this network with this strength is equivalent to take the 
one-mode approach to the bipartite structure with weighted links, 
whose value is given by $w_M$. More specifically, by imposing a minimum 
threshold value for $w_M$, we can unveil the fractality that was hidden below 
many links that were taken into account in the original network. In other words, 
our network is no longer small-world, but it is self-similar. This behavior 
is reminiscent of the transition calculated in the RG analysis of \cite{rozenfeld10}.
Given that the self-similarity can be seen already at $w_M=2$, the full connection 
mode considers a great number of links that hide fractality and make the network 
look like a small-world case. Another way to look at this fact is that 
connecting all actors in a movie is equivalent to put in equal footing 
strong, long lived collaborations (like that between Katherine Hepburn 
and Spencer Tracy, one of the most famous couples in film history, 
which were cast in $6$ movies together) and weak ones 
(like those of the 1956 movie `Around the world in eighty days' 
in which 1297 distinct actors are credited, which generates 
$\approx1.68\times10^6$ links). Clearly, the first ones, which are more 
specialized, are responsible for the modular character of the network.

\section*{Results}
\begin{figure}[!ht]
\begin{center}
\centerline{\resizebox{8.0cm}{!} { \includegraphics[angle=-90]{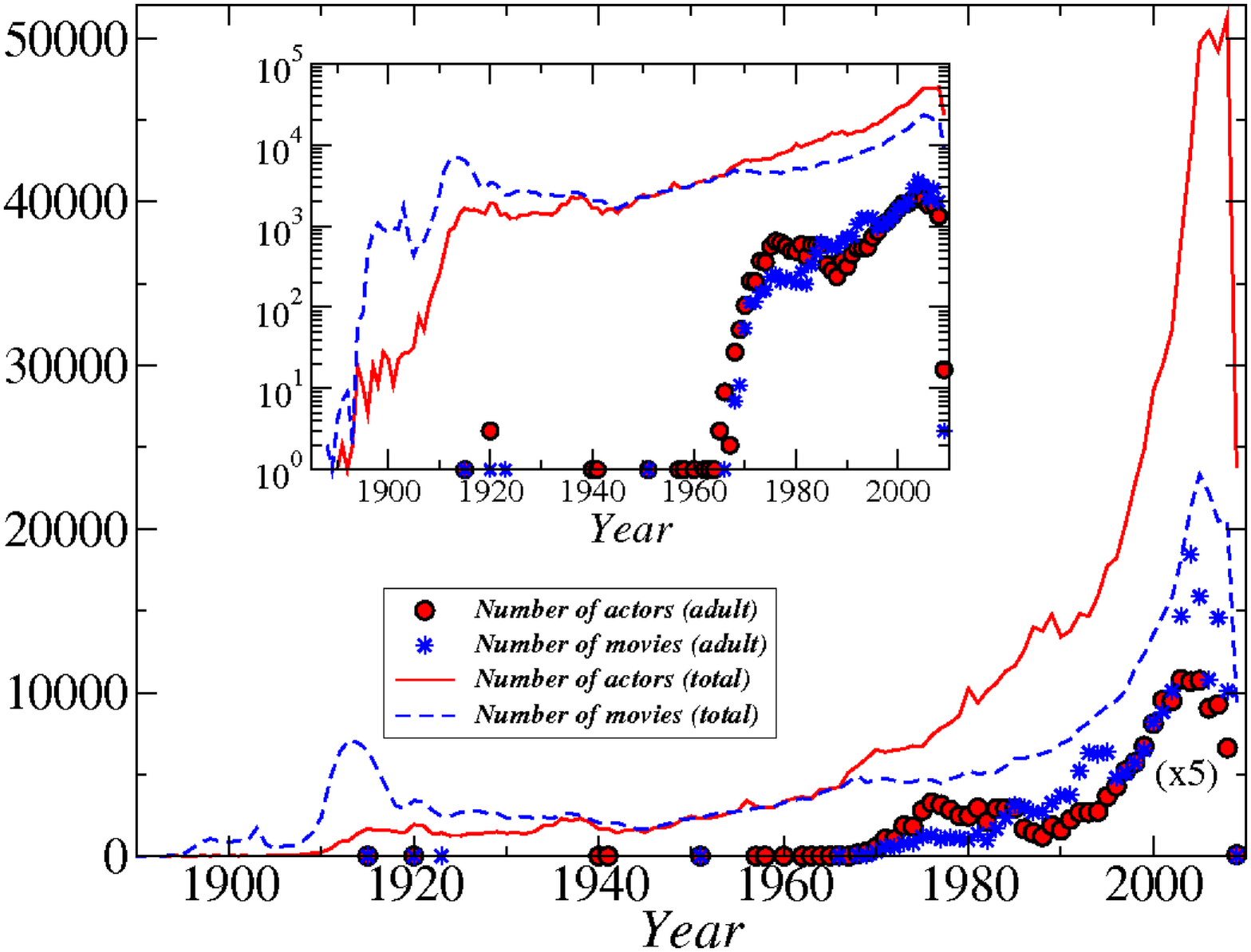}}}
\end{center}
\caption{
{\bf Number of actors per year and number of movies per year for the entire IMDB database (lines) and for the adult IMDB database (symbols)}. In the plot the adult IMDB points have been multiplied by a factor of 5 for clarity. Inset: the same numbers on a logarithmic y-scale.}
\label{fig_peryear}
\end{figure}

The time evolution of the movie industry is shown in figure~\ref{fig_peryear}.
We associate each actor with a unique year, which is the average year of
all the movies at which she/he participated. In this way, we can approximate the
time that an actor was more active, although obviously she/he may have starred
for decades.
The number of movies and the number of actors increase monotonically
with time as expected, both in the IMDB dataset and in the adult IMDB dataset. 

The adult movie industry has a much more recent history, since movies started appearing in bulk only after circa 1990,
and the increase in their production rate in these years is much higher than for all movies.
The network representation, thus, between adult actors is more robust since it
includes a total of only 20-30 years, compared to more than 100 of the typical
IMDB database.

\begin{figure}[!ht]
\begin{center}
\centerline{ \resizebox{8cm}{!} { \includegraphics[angle=-90]{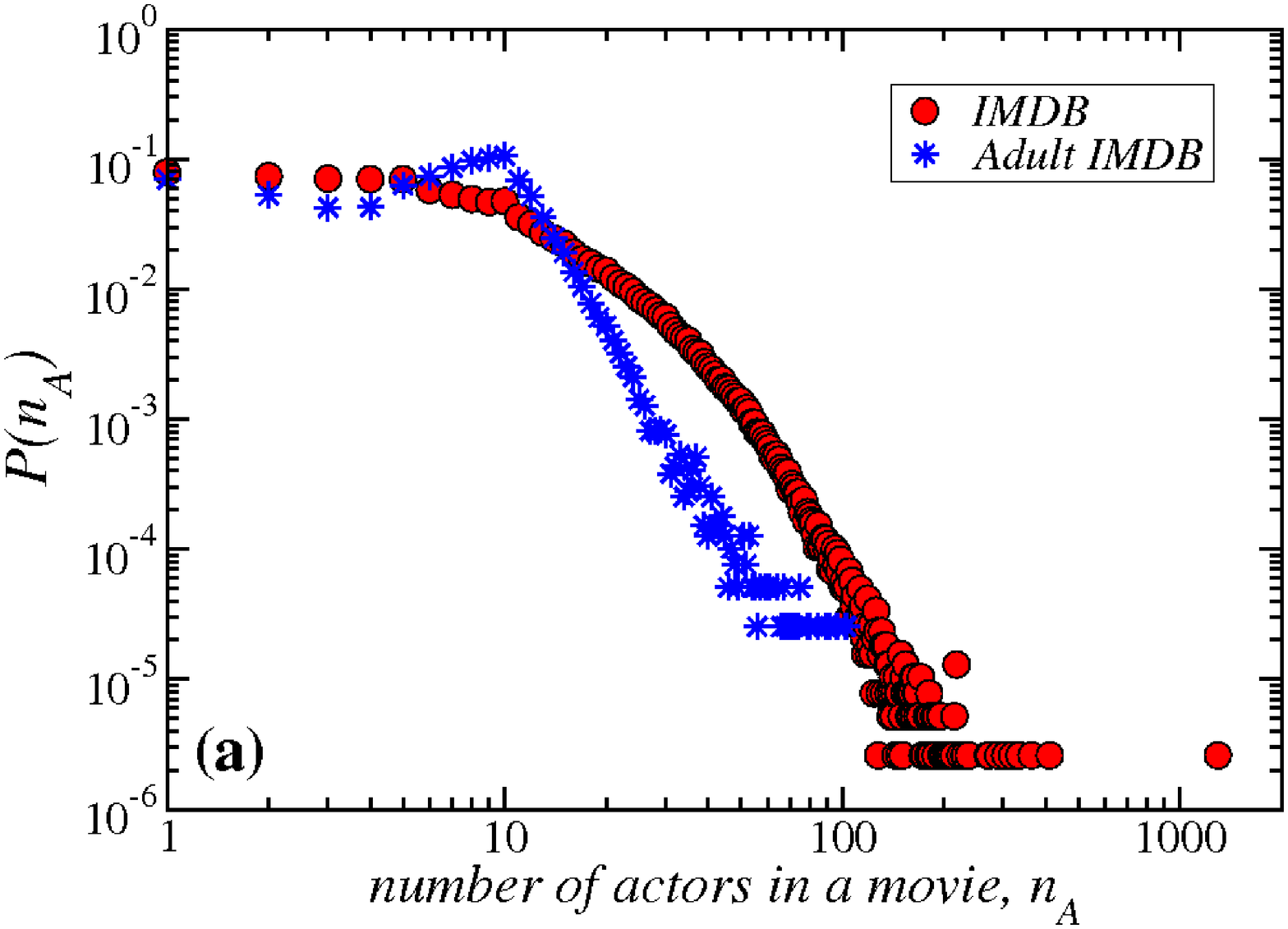}}
\resizebox{8cm}{!} { \includegraphics[angle=-90]{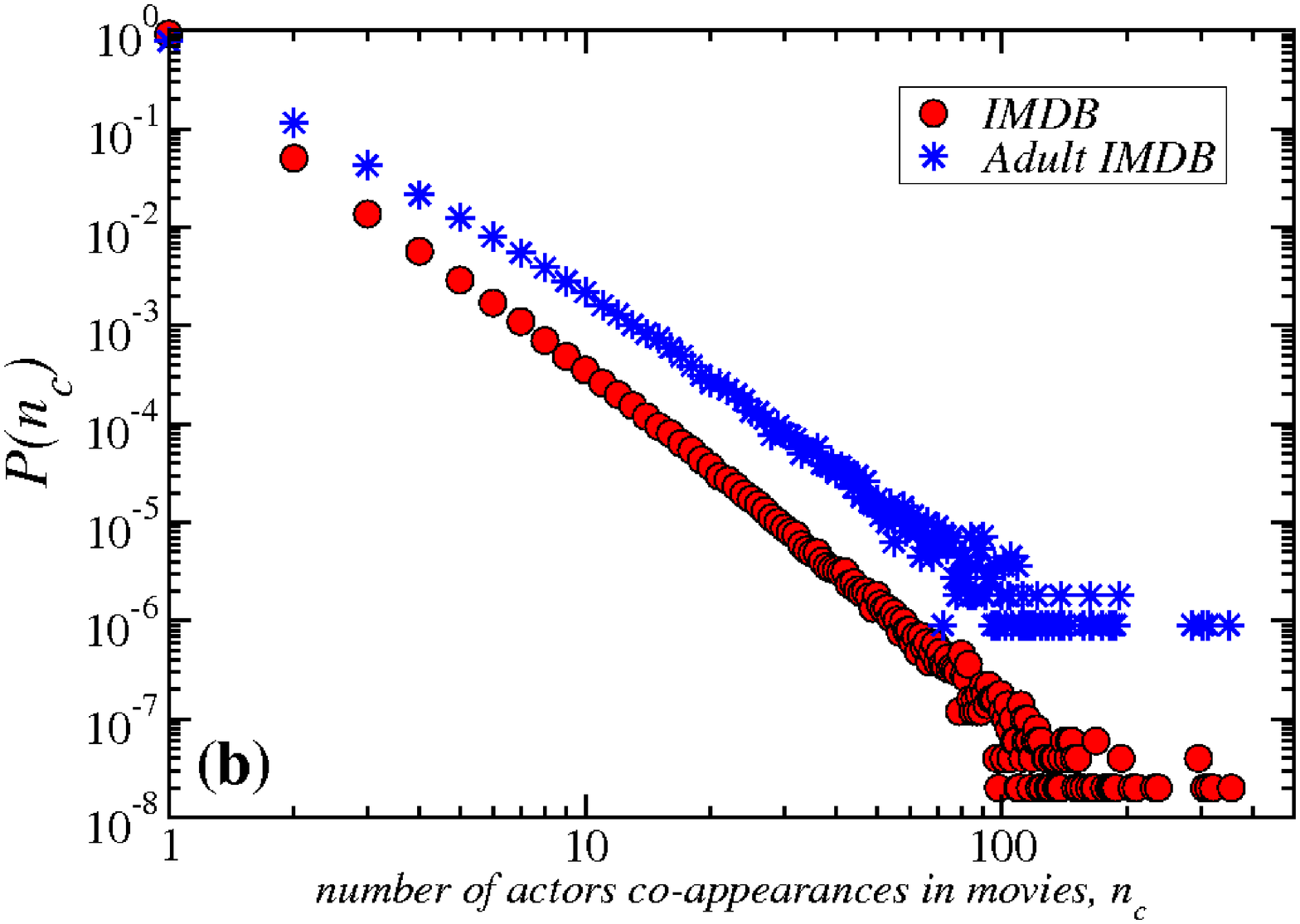}}}
\end{center}
\caption{
{\bf Comparison of the IMDB network to the adult IMDB network.} a) Probability distribution that a number $n_A$ of actors participated in a movie (as reported in IMDB). b) Distribution of the links weights, or equivalently, the probability that two actors have participated in $n_c$ movies together.}
\label{fig_comparison}
\end{figure}

Initially we connected all the actors that participated in the same adult 
movie. The resulting largest connected component comprises 
44719 actors/actresses, out of a total of 39397 movies. The average degree is 
$\langle k \rangle=46$, which is quite high
but is still significantly smaller than the corresponding average for the entire
IMDB network (where $\langle k \rangle=98$). Both the number of reported actors
per movie and the probability that two actors participated in $n_c$ common movies
decay as power laws for both networks, with similar exponents (figure~\ref{fig_comparison}).
This points to the inhomogeneous character of the network, both in terms of diverging
number of actors in a movie and in terms of how many times two actors collaborated
with each other in their careers, a number which can also vary significantly from 1 to 350 times.

\begin{figure}[!ht]
\begin{center}
\centerline{ \resizebox{8.0cm}{!} { \includegraphics{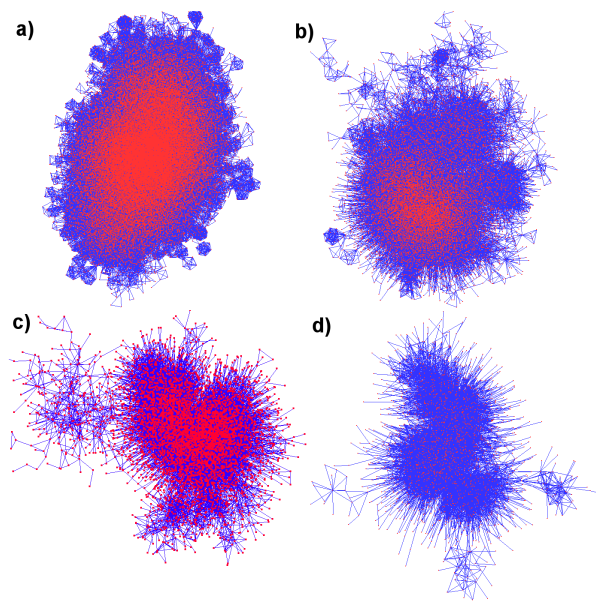}}}
\end{center}
\caption{
{\bf Snapshots of the `adult IMDB' network, where two actors are connected if they have co-starred in at least a) $w_M=1$, b) $w_M=2$, c) $w_M=4$, or d) $w_M=8$ movies, respectively.} Only the largest connected component is shown, and the corresponding network sizes are $N=47719$, 14444, 5315, and 2100 actors, respectively. 
}
\label{fig_snapshot}
\end{figure}

Naturally, the resulting collaboration network is weighted. Every link can be considered
as carrying a different strength, depending on the number of movies in which 
two 
actors have participated. For example, the strongest link appears between two 
actors who have co-starred
in 350 movies and obviously this connection is more important than between two
actors with just one common appearance. Thus, we filter our network by imposing
a threshold $w_M$, so that two actors are connected only if they have collaborated in
at least $w_M$ movies, with a larger $w_M$ value denoting a stronger link.
Snapshots of the largest connected component of networks with varying $w_M$ values are shown in figure~\ref{fig_snapshot}.
It is clear from the plots that for $w_M=1$ there are many fully-connected modules
and some of them appear largely isolated, since very few of the actors in that movie
participated in any other movies. However, as we increase $w_M$ the network tends
towards a more tree-like structure with less loops.

\begin{figure}[!ht]
\begin{center}
\centerline{ \resizebox{10.0cm}{!} { \includegraphics[angle=-90]{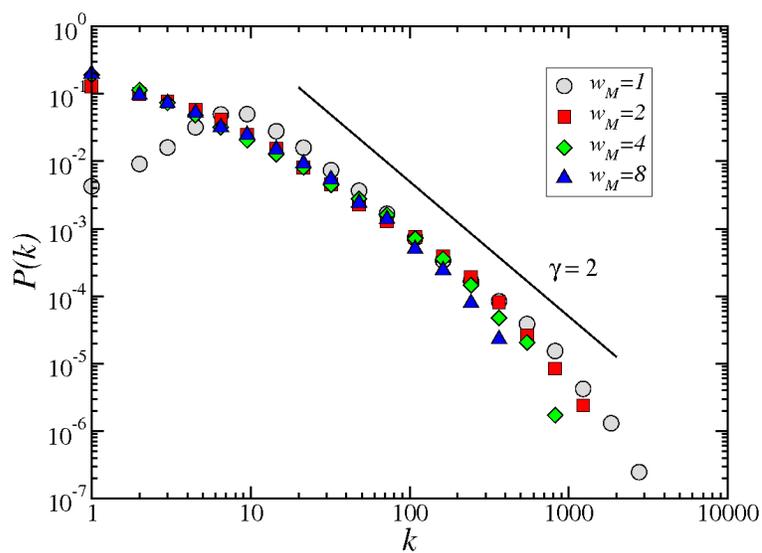}}}
\end{center}
\caption{
{\bf Degree distribution of the adult IMDB network, for different connection strengths, $w_M=1$, 2, 4, or 8 movies, respectively.}The solid line has a slope, the exponent of the power law degree distribution, $\gamma=2$ (see text). 
}
\label{fig_Pk}
\end{figure}

All of these networks exhibit a scale-free degree distribution, where asymptotically
the number of links $k$ scales as a power law, i.e. $P(k)\sim k^{-\gamma}$, where $\gamma$
is the degree exponent with a value around $\gamma=2$ (figure~\ref{fig_Pk}). The unweighted
network, though, behaves differently as compared to the other networks at small $k$.
When $w_M=1$, there are very few nodes with 1 or 2 connections. This percentage
increases with $k$ until we reach a maximum at $k=10$, and afterwards decays as a power-law.
In contrast, the decrease in the degree distribution is monotonic for any $w_M>1$
and the distribution remains almost invariant, independently of the $w_M$ value.
The shifted peak for $w_M=1$ is a result of connecting all the actors in a movie
with each other, which makes it rare for an actor to be related with only a small number
of other actors. In contrast, the shape of the distribution follows a more typical pattern
when we retain stronger connections only, even for $w_M=2$.

\begin{figure}[!ht]
\begin{center}
\centerline{ \resizebox{10.0cm}{!} { \includegraphics[angle=-90]{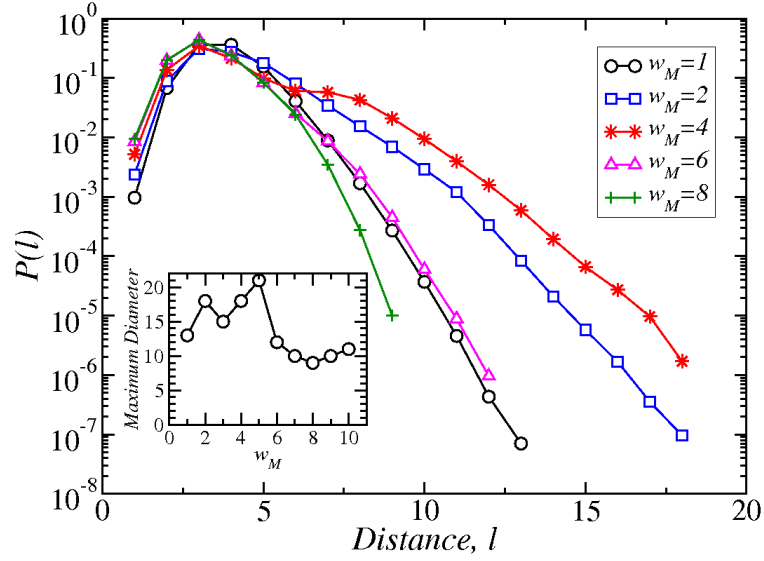}}}
\end{center}
\caption{
{\bf Probability distribution of the shortest paths in networks with varying weight threshold $w_M$.}
 Inset: Maximum diameter in the network as a function of $w_M$. 
}
\label{fig_ldistr}
\end{figure}

The process of increasing $w_M$ corresponds to diluting the network through the
removal of a large number of links.
As a result, we expect an increase of the network diameter
(defined as the largest distance between all possible shortest
paths) since we destroy many of the existing paths.
However, at the same time many nodes are removed from the largest cluster as well, and the
dependence of the diameter on $w_M$ cannot be easily predicted. In figure~\ref{fig_ldistr}
we can see that the interplay between removing links and nodes results in a more
complicated behaviors. As we increase $w_M$, longer distances appear in the network
as a result of removing more links than nodes. At $w_M=5$, though, the maximum diameter
drops significantly, because now the network size has also decreased a lot.
The main peak of the distribution remains constant at distances $\ell$ between 
$3$ and $4$. The 
tail of the distribution, though, becomes initially wider with increasing $w_M$,
but after $w_M=5$ the tail shifts back towards smaller values.

\begin{figure}[!ht]
\begin{center}
\centerline{ \resizebox{10.0cm}{!} { \includegraphics[angle=-90]{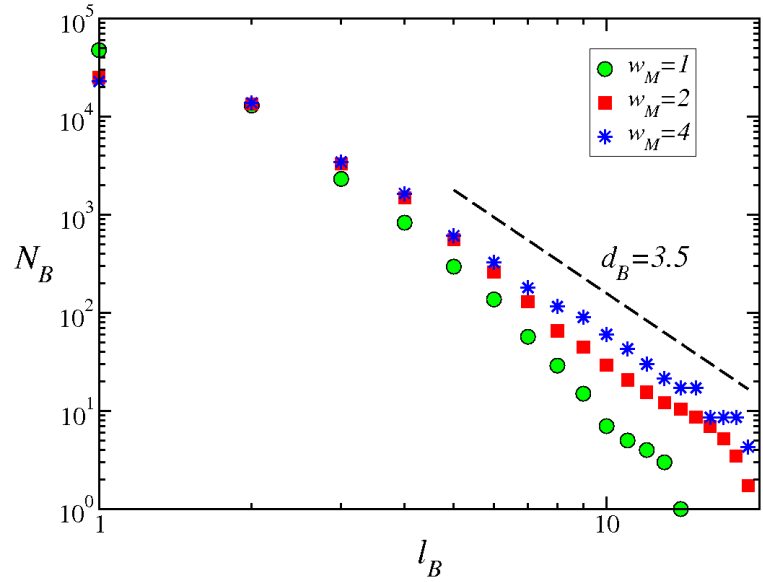}}}
\end{center}
\caption{
{\bf Scaling of the number of boxes $N_B$ as a function of the maximum box diameter $\ell_B$, for the adult IMDB network and for different $w_M$ values.} Except for $w_M=1$, the other networks have a fractal dimension, shown by the dashed line with slope, $d_B=3.5$.
}
\label{fig_db}
\end{figure}

The IMDB network has been shown to possess fractal characteristics and
we have found the same property for the adult IMDB network, as well. 
We used the box-covering technique \cite{song07} to perform this analysis.
The number of boxes $N_B$ that
are needed to optimally cover the network scales with the maximum box diameter $\ell_B$
as a power law (figure~\ref{fig_db}). The scaling is much more prominent
for values of $w_M>1$, indicating that connecting all the actors in a movie ($w_M=1$)
results in an over-connected network which may alter the apparent behavior of many measured 
properties. The scaling is much smoother for the presented values of $w_M=2$ or 4,
with a fractal exponent $d_B=3.5$.



\begin{figure}[!ht]
\begin{center}
\centerline{ \resizebox{10.0cm}{!} { \includegraphics[angle=-90]{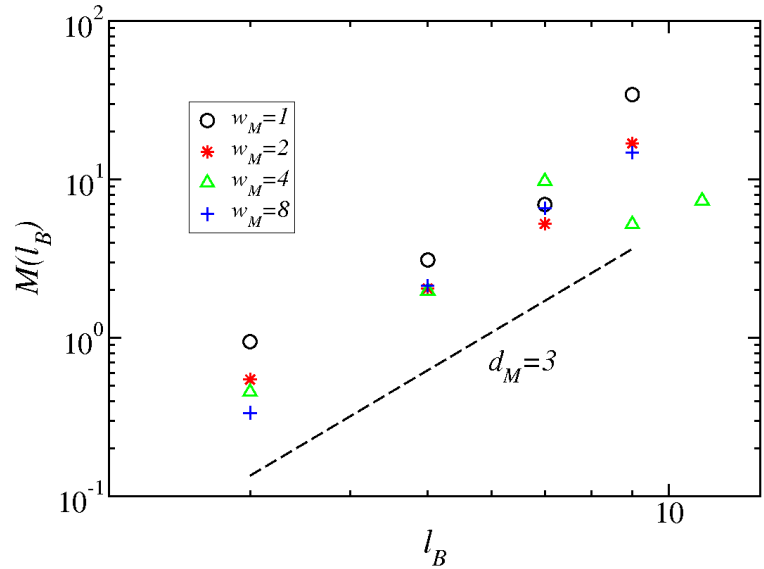}}}
\end{center}
\caption{
{\bf Scaling of the modularity $M$ as a function of the maximum box diameter $\ell_B$, for the adult IMDB network and for different $w_M$ values.} All networks scale with $\ell_B$, with a slope close to $d_M=3$, as shown by the dashed line with the same slope value.
}
\label{fig_modularity}
\end{figure}

Similarly, we have measured the degree of modularity in these networks, and how 
this modularity scales with the box size $\ell_B$ (figure~\ref{fig_modularity}).
The process is as follows: First, we fix the value of $\ell_B$ and we
cover the network with the minimum possible number of boxes, where every two nodes
in the box are in a distance smaller than $\ell_B$. These boxes can then be considered as
modules, in the sense that all nodes in a box have to be close to each other.
The level of modularity remains then to be shown, and following \cite{galvao10,gallos12},
we can define the quantity
\begin{equation}
M(\ell_B) = \frac{1}{N_B} \sum_{i=1}^{N_B} \frac{L_i^{\rm in}}{L_i^{\rm out}} \,,
\end{equation}
where $L_i^{\rm in}$ and $L_i^{\rm out}$ represent the number of links that start in a given
box $i$ and end either within or outside $i$, respectively. Large values of $M$ correspond,
thus, to a higher degree of modularity.
The meaning of $M(\ell_B)$, as defined above, does not correspond directly to other measures
of modularity in the literature, designed to detect modules in a given network, independently of the module
size. Here, the value of $M$ varies with $\ell_B$,
so that we can detect the dependence of modularity on different length scales, or equivalently
how the modules themselves are organized into larger modules that enhance or decrease modularity,
and so on. This information can be detected more reliably through the modularity exponent $d_M$, which
we have defined in terms of the following power law relation:
\begin{equation}
M(\ell_B) \sim \ell_B^{d_M} \,.
\end{equation}
The value of $d_M=1$ represents the borderline case that separates modular ($d_M>1$)
from random non-modular ($d_M<1$) networks. For example, in a lattice structure the value of $d_M$ is
exactly equal to $1$, while a randomly rewired network has $d_M=0$. 

It is not difficult to understand how these two values arise. Let us, first, 
consider a regular network optimally tiled with boxes with large 
$\ell_B$. The number of links that terminate outside a box originate only in 
nodes positioned along the boxes' surface, given the regularity of the network, i.e, there are no long 
range links. The number of surface links is proportional to the surface of the 
box, 
$L_{i}^{\rm out}\sim\ell_B^{d-1}$, where $d$ is the spatial dimension of the box. 
On the other hand, links that terminate inside this box are clearly 
related to the number of bulk nodes, i.e., nodes that are not at the surface of 
the box. The number of bulk nodes is proportional to the boxes' volume, 
$L_{i}^{\rm in}\sim\ell_B^d$. Finally, the ratio of these two quantities yields 
$d_M=1$.

This continuum argument cannot be used for a randomly rewired network because 
there may be links that connect nodes which are not first neighbors in the 
original regular lattice. As a consequence of this fact, a link that ends 
outside a box can start at any node within this box. Following the same reasoning, we 
conclude that a link that ends within a box can be originated in any node 
in this box. As we increase $\ell_B$, new nodes will be added, whose links are 
random as well. Therefore, a box with a distinct size bears a link 
configuration which is not substantially different from a box with a smaller 
$\ell_B$. We conclude, then, 
that the modularity, i.e., the number of links that terminate within or 
outside a box, is not affected by the boxes' size. In other words, $d_M=0$.

For the adult IMDB
network we find an approximate exponent of $d_M=3$ for all different values of $w_M$, which is an indication of a highly modular network.

The standard method of calculating fractality through the scaling of $N_B$ vs $\ell_B$
may sometimes yield ambiguous results, due to the relatively short range of $\ell_B$.
The renormalization group (RG) analysis \cite{rozenfeld10}, though, can provide a more reliable picture,
and can highlight the self-similarity of a network under continuous scale transformations.
The RG analysis is particularly useful in revealing long-range links (shortcuts) in the structure,
since under renormalization these links will either persist or disappear.

In the case of the adult IMDB network, our finding of fractality and modularity
is a strong indication that there are not many such long-range links in the network.
This is an important concept that can be shown to determine the efficiency in navigability or the efficient
information transfer in these networks \cite{gallos12,kleinberg00,li10}. In order to employ the RG analysis for the study of shortcuts, 
we again use the box-covering technique \cite{song07}.

\begin{figure}[!ht]
\begin{center}
\centerline{ \resizebox{14.0cm}{!} { \includegraphics{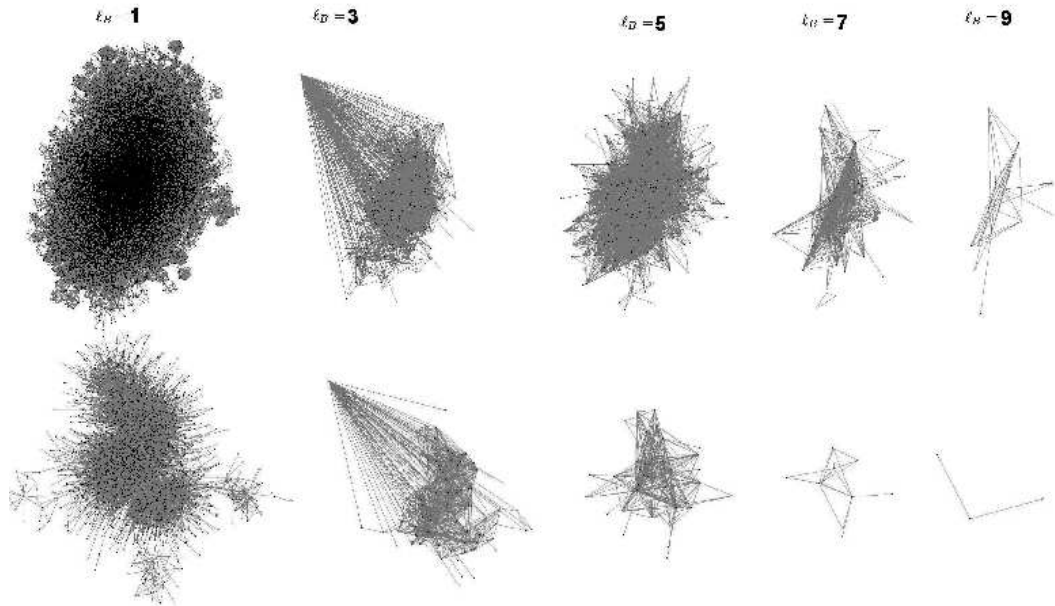}}}
\end{center}
\caption{
{\bf Networks resulting from the box-covering procedure at different $\ell_B$ values.} Top row: Adult IMDB network at $w_M=1$. Bottom row: Adult IMDB network at $w_M=8$.
}
\label{fig_renorm}
\end{figure}

This renormalization analysis can provide very accurate information on the structure of self-similar objects.
Snapshots of the resulting networks at different $\ell_B$ values are
shown in figure~\ref{fig_renorm}.

\begin{figure}[!ht]
\begin{center}
\centerline{ \resizebox{10.0cm}{!} { \includegraphics[angle=-90]{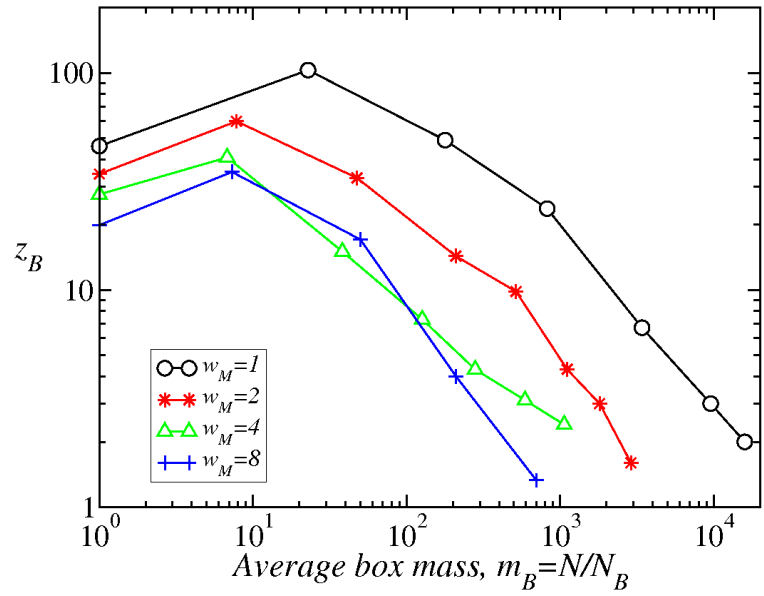}}}
\end{center}
\caption{
{\bf Variation of the average degree in the IMDB network for different $w_M$ values, as a function of the average box mass $m_B=N/N_B$.}}
\label{fig_zb}
\end{figure}

In the case of the adult IMDB network, we find that the exponent $\lambda$, see equation (\ref{rg_eq}), is always negative
(figure~\ref{fig_zb}). This shows that nodes in the network tend to remain connected within their close
neighborhood, and there are very few links that span different areas of the network, i.e., under successive renormalization steps, shortcuts in the network tend to disappear and we are left with a fractal structure. In this figure, we see that as we increase $w_N$, the curves become independent on this quantity. This implies that for a high enough link strength, it has the same average distribution of links.




\section*{Methods}
The box-covering technique \cite{song06} consists in covering the network with boxes of size $\ell_B$. 
This means that within a box, there will be only vertices that are separated from each other by, at most, 
$\ell_B-1$ links. 
These boxes then become the super-nodes, which are connected with each other if there is at least one 
connection between nodes within these boxes. This yields a network representation at a different scale, 
which can in turn be covered, again, with boxes of size $\ell_B$. This new network is renormalized 
again, and the process repeats with progressively coarser covering.\\

\section*{Conclusions}
We have studied a subset of the IMDB network, the adult IMDB, as an one-mode 
projection of the original bipartite IMDB network. Therefore, we were able to remove many pathologies of the full IMDB, such as 
actors with long-spanning careers, and have taken a network more focused in time.

We have also imposed a minimum link strength threshold, $w_M$, such that our 
network is composed of vertices that have at least this amount of links 
between them. We showed that the adult IMDB is a small-world network with an 
exponent for the distribution of links $\gamma=2$, if $w_M\geq2$. 
We have seen that this procedure removed a large number of 
links between actors. These links changed the distribution of links in a way 
that it was more likely to find an actor with $k=10$ than one with $k=2$. Since 
the average degree is larger for the network at $w_N=1$, it is easy to cross 
the whole network in a smaller number of steps than to cross the same network built at $w_N>1$. The unweighted network is very similar to a typical small-world case, while those at larger $w_N$ are closer to a fractal. In conclusion, we 
see that the parameter $w_N$ controls a transition from a small-world to a fractal network, in much the same way as wa shown in \cite{rozenfeld10}

We confirmed these conclusions with measurements of the modularity and with a renormalization group analysis of the network. We obtained that the modularity 
scales with the largest box diameter with an exponent of $d_B=3$, which renders 
it a highly modular network. In addition to an efficient community separation, 
we were able to show that this is not a typical small-world network. This finding was based on a renormalization group 
analysis, which we have shown to be a robust method for detecting fractality. This 
analysis revealed that the difference between the average number of nodes in the renormalized 
and the original network scale with $\lambda<0$, a signature for large-world behavior.
The highly modular fractal characteristics found here are expected to naturally emerge in many 
social networks, but in general they are hard to accurately detect in real-world networks. Ultimately, we feel that our weighting procedure constitutes a 
reliable way of unveiling fractality out of a small-world collaboration 
network, since the information lost by choosing larger $w_M$ values 
actually hides the fractal network.

\section*{Acknowledgments}
We acknowledge support from the Brazilian agencies CNPq, CAPES, and FUNCAP, the FUNCAP/Cnpq Pronex grant, and the National Institute of Science and Technology for Complex Systems in Brazil for financial support.

\bibliography{potiguar}

\section*{Figures}

\end{document}